\documentclass[a4paper,11pt]{article}
\usepackage{pos}
\usepackage{siunitx}
\manuallySeparateAuthors
\renewcommand{\logo}{\relax} 

\title{ALICE Upgrades}

\author*[a]{Felix Reidt}
\author[]{ on behalf of the ALICE collaboration}

\affiliation[a]{CERN,\\
  Esplanade des Particules 1, 1217 Meyrin, Switzerland}

\emailAdd{felix.reidt@cern.ch}
\abstract{

The ALICE collaboration prepares multiple upgrades to further extend the reach of heavy-ion physics at the LHC. For LHC Run 4 (2030--2033), a Forward Calorimeter (FoCal) system combines a high-granularity electromagnetic silicon-tungsten calorimeter with a conventional hadronic calorimeter leading to an excellent two-shower separation for neutral pion reconstruction and photon isolation. Direct photon measurements with FoCal will uniquely constrain the low-\(x\) gluon structure of protons and nuclei via forward measurements of direct photons. \\
Additionally, ALICE will employ bent, wafer-scale pixel sensors achieving truly cylindrical tracking layers with very low material budget. These layers will replace the three innermost layers of the Inner Tracking System resulting in an improvement in pointing resolution allowing new measurements of heavy-flavour hadrons and dielectrons.\\

For Run 5 and beyond, a next-generation detector system, ALICE 3, has been conceived to gain unique access to the interaction and thermalization of heavy flavour probes in the QGP as well as to the thermal radiation carrying information about the temperature and the restoration of chiral symmetry.  At its core, it combines a high-resolution vertex detector with a large-acceptance silicon pixel tracker. For the identification of particles, a combination of a time-of-flight system, a ring-imaging Cherenkov detector, an electromagnetic calorimeter, a muon identifier, and a dedicated forward detector for ultra-soft photons, are envisaged. 
}

\FullConference{%
12th Large Hadron Collider Physics Conference -- LHCP 2024\\
3-8 June 2024\\
Boston, MA, US
}

\begin{document}
\maketitle

\section{Introduction}

ALICE is designed to study the quark--gluon plasma produced in heavy-ion collisions at the LHC. Two main physics items are driving its upgrade strategy:
\begin{itemize}
  \item Studies of heavy flavour transport and hadronization in the medium via differential measurements of hadron production down to vanishing transverse momentum;
  \item Measuring electromagnetic radiation from the medium through dileptons below the \(J/\psi\)-mass down to zero transverse momentum aiming to map out the evolution of the collision.
\end{itemize}
These measurements are carried out best with a high-granularity, low-mass detector with continuous readout in order to access signals with very low signal-over-background ratio. Additionally, particle identification of hadrons, leptons and photons over a broad acceptance is crucial for the physics programme. This strategy is also strongly beneficial for a set of other physics areas, ranging from the study of collective effects, hyperons and (hyper)nuclei, to (heavy flavour) jet substructure and ultra-peripheral collisions.

As part of this strategy ALICE pursues several upgrades. During the Long Shutdown 3 (LS3) of the LHC planned for the years 2026 to 2030, a new Forward Calorimeter (FoCal) (cf.~Sec.~\ref{sec:FoCal})~\cite{CERN-LHCC-2024-003} and the Inner Tracking System 3 (ITS3) (cf.~Sec.~\ref{sec:ITS3})~\cite{CERN-LHCC-2024-004} will be installed. For Long Shutdown 4 (LS4) in 2034 and 2035, the ALICE collaboration plans to replace the existing experimental apparatus by an entirely new one, called ALICE 3 (cf.~Sec.~\ref{sec:ALICE3})~\cite{ALICE:2022wwr}.

\section{Forward Calorimeter (FoCal)}\label{sec:FoCal}
The Forward Calorimeter (FoCal) upgrade~\cite{CERN-LHCC-2024-003} primarily targets the search for evidence of non-linear Parton-Distribution function evolution in Quantum Chromodynamics (QCD) in nucleons and nuclei at low Bjorken-\(x\) down to \(10^{-6}\).

\begin{figure}[b]
  \centering
  \includegraphics[width=0.45\textwidth]{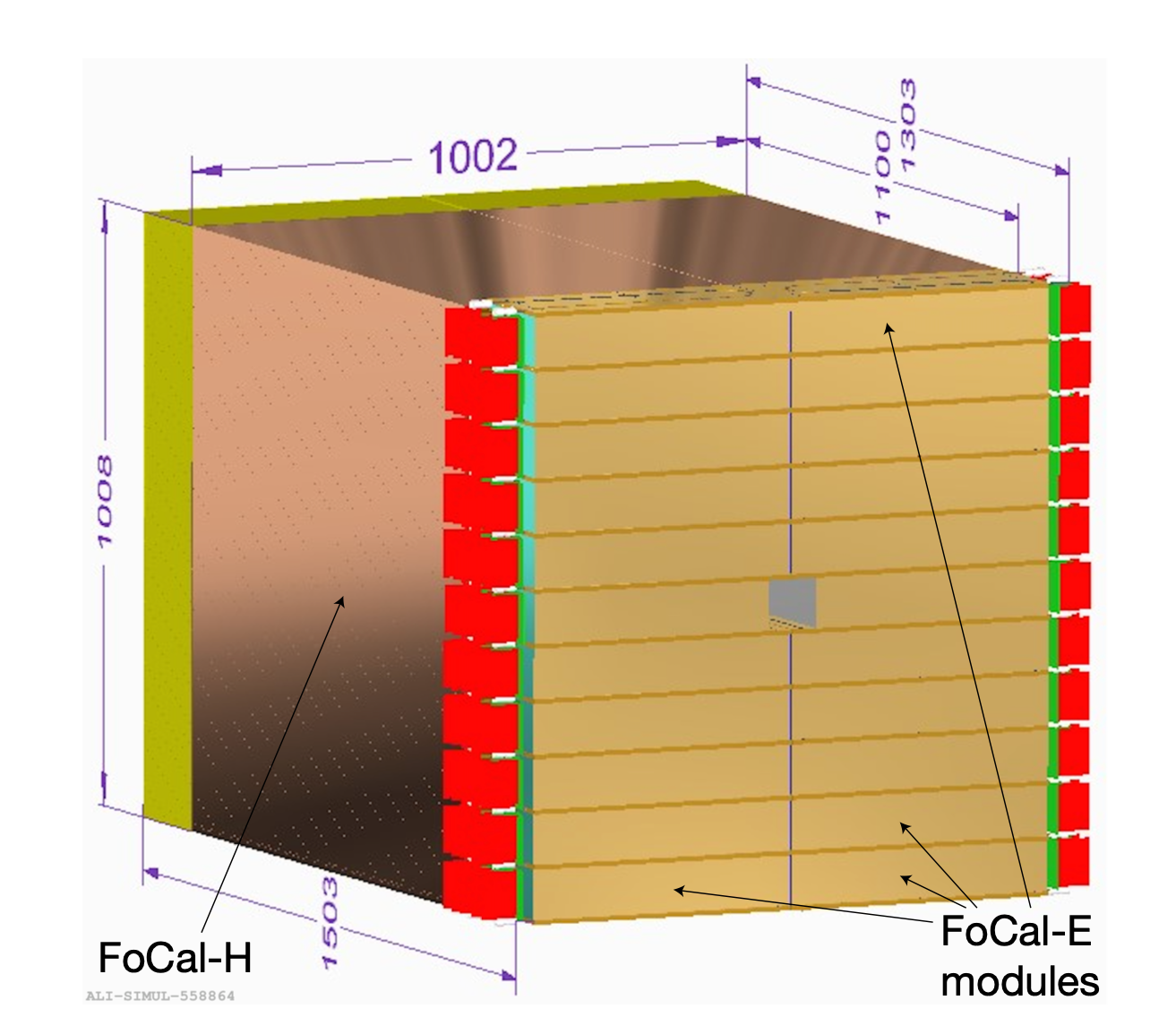}
  \includegraphics[width=0.45\textwidth]{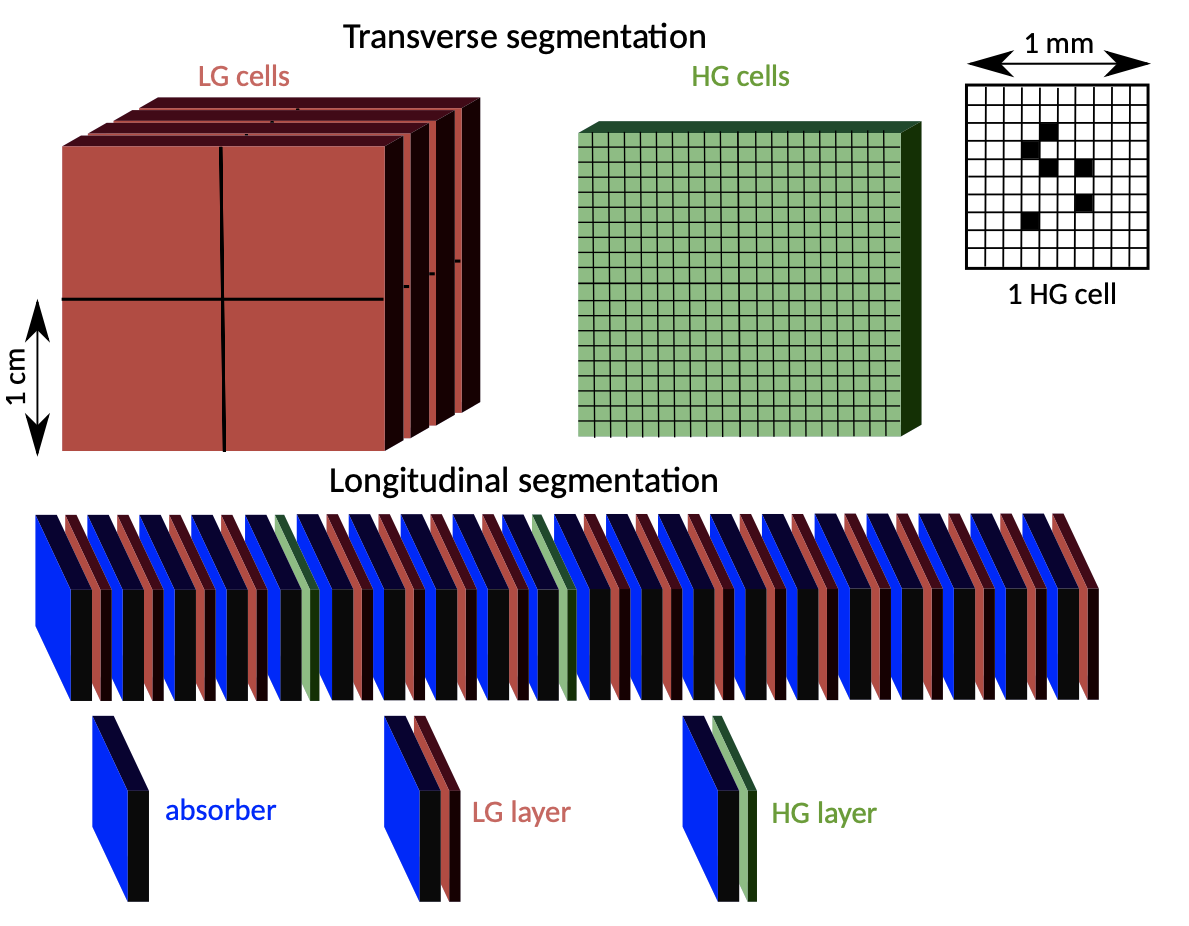}
  \caption{\textbf{Left:} sketch of FoCal, dimensions given in millimetres; \textbf{right}: sketch of FoCal-E, showing 20 layers of \qty{3.5}{\milli\meter} of W plus Si sensors. There are 18 layers consisting of pad sensors ('LG') and two layers of pixel sensors in positions five and ten, both taken from~\cite{CERN-LHCC-2024-004}.\label{fig:focal}}
\end{figure}

\begin{figure}
  \includegraphics[width=0.95\textwidth]{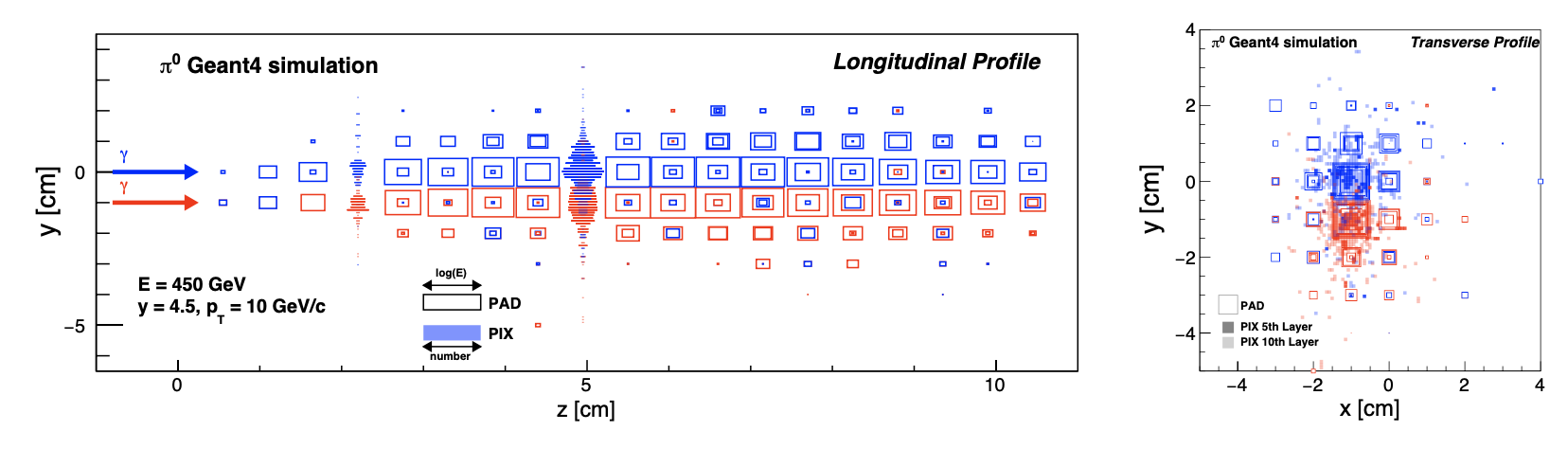}
  \caption{Longitudinal (left) and transverse (right) profile of two showers originating from high-energy photons displaced by \qty{1}{cm} on the surface of FoCal-E. The energy deposits in the pad and pixel sensors are colour-coded in red and blue, respectively, taken from~\cite{CERN-LHCC-2024-004}.\label{fig:focalShowerProfile}}
\end{figure}

FoCal (cf.~Fig.~\ref{fig:focal}) consists of an electromagnetic and hadronic calorimeter with a coverage of \(3.2<\eta<5.8\) in pseudorapidity which will be used in a multi-messenger approach.
\begin{description}
  \item[FoCal-E]: a compact silicon-tungsten sampling electromagnetic calorimeter featuring Low-Gran\-u\-lar\-i\-ty (LG) pad layers (\qtyproduct{1 x 1}{\centi\meter}) and High-Granularity (HG) pixel layers (\qtyproduct{30 x 30}{\micro\meter}) (cf.~Fig.~\ref{fig:focal}, right) providing high spatial resolution allowing one to discriminate between isolated photons and photon pairs from hadron decays as depicted in Fig.~\ref{fig:focalShowerProfile}.
  \item[FoCal-H]: a hadronic calorimeter constructed from copper capillary tubes filled with scintillating fibres providing photon isolation, energy and jet measurements.
\end{description}

FoCal will provide forward measurements of direct photons, neutral mesons, vector mesons, jets, Z-bosons, and their correlations in hadronic and ultra-peripheral p--Pb and Pb--Pb collisions~\cite{ALICE-PUBLIC-2023-001}. The expected physics performance is summarized in~\cite{ALICE-PUBLIC-2023-004}.

\begin{figure}
  \centering
  \includegraphics[width=0.8\textwidth]{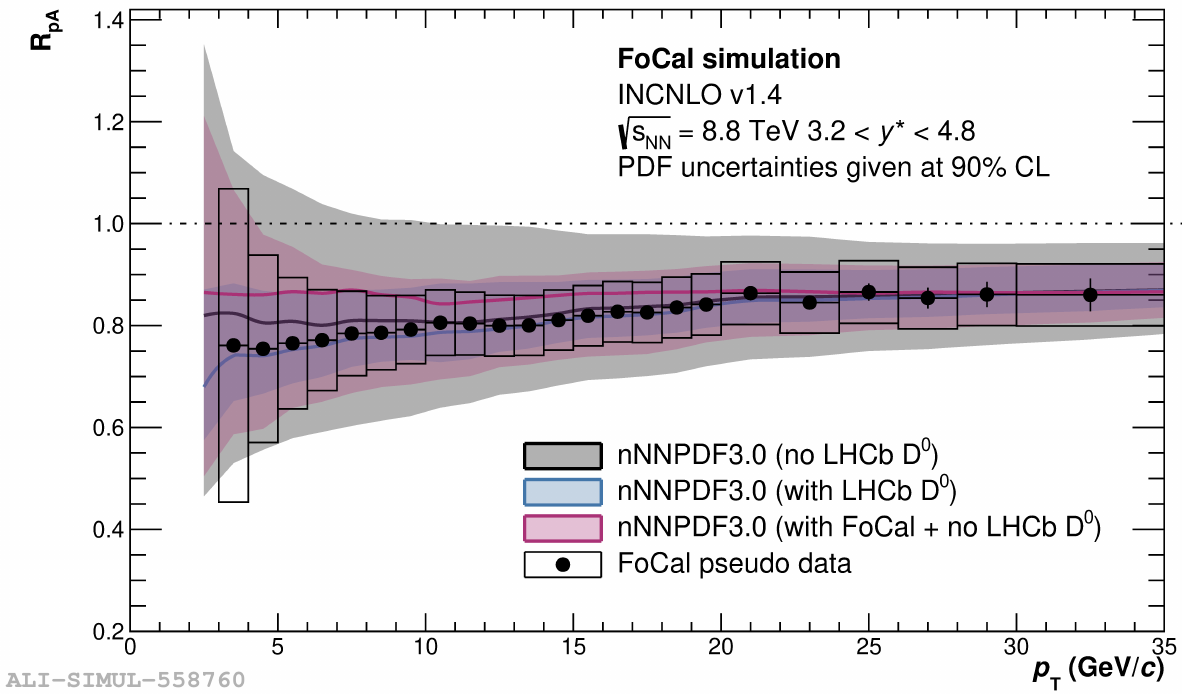}
  \caption{Distribution of the nuclear modification factor $R_\mathrm{pPb}$s, for prompt photons obtained from FoCal pseudodata and QCD calculations using nNNPDF3.0 with and without LHCb D-meson measurements, taken from~\cite{ALICE-PUBLIC-2023-004}.\label{fig:focalRpA}}
\end{figure}

A key FoCal measurement constraining nuclear Parton Distribution Functions (nPDFs) is the $R_\mathrm{pPb}$, the ratio of inclusive prompt-photon yields measured in p--Pb and pp collisions. Figure~\ref{fig:focalRpA} shows the calculated distribution of the isolated prompt photon $R_\mathrm{pPb}$ in comparison to QCD calculations performed with INCNLO incorporating nNNPDF3.0 without constraint from LHCb D-mesons (black line, corresponding band indicating uncertainties), together with reweighed distributions incorporating LHCb D-mesons~\cite{Aaij2017} (blue) or together with FoCal prompt-photon pseudo-data (red). Both the inclusion of LHCb D-mesons and the FoCal prompt photon measurement lead to a sizeable reduction of the nPDF uncertainties. The two probes are complementary as photon measurements are insensitive to final-state effects.

\begin{figure}
  \centering
  \includegraphics[width=0.46\textwidth]{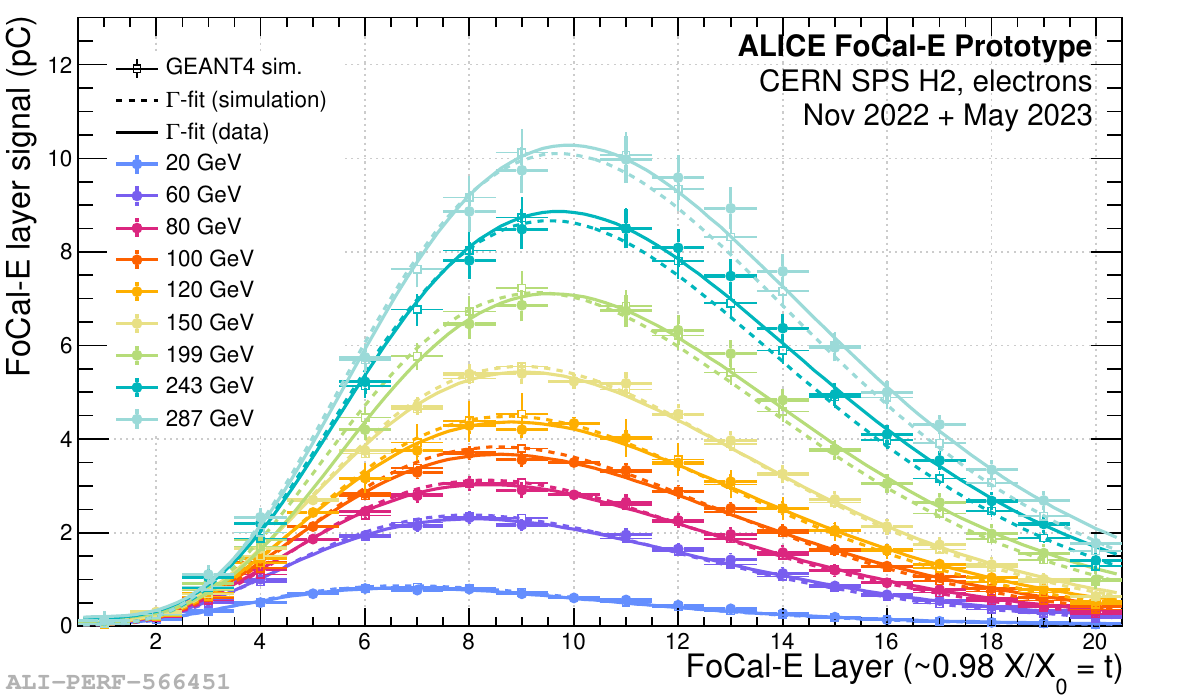}
  \includegraphics[width=0.53\textwidth]{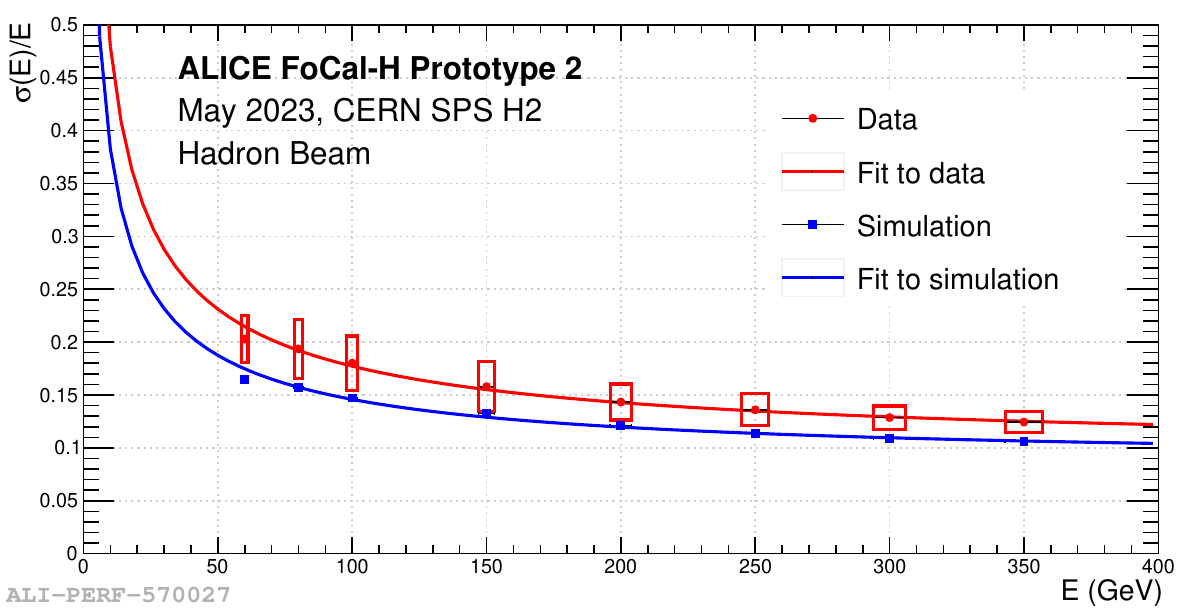}
\caption{\textbf{Left:} longitudinal shower profiles for \qtyrange{20}{300}{GeV} electrons compared to Geant4 simulations and fitted with a \(\Gamma\)-distribution; \textbf{right:} energy resolution of the FoCal-H prototype as function of the hadron beam energy compared to simulations and respective fits, taken from~\cite{Aehle_2024}.\label{fig:focalTb}}
\end{figure}

The performance of full-length prototype of FoCal has been studied using test beam at the CERN PS and SPS facilities with hadron beams up to energies of \qty{350}{GeV}, and electron beams up to \qty{300}{GeV}~\cite{Aehle_2024}. Results from this campaign demonstrate the capabilities of FoCal. For the FoCal-E pad sensors, a good agreement between data and simulations for longitudinal shower profiles for \qtyrange{20}{300}{GeV} electrons is shown in Fig.~\ref{fig:focalTb} (left). The relative energy resolution was found to be lower than 3\% at energies larger than \qty{100}{GeV}. The transverse shower profiles, evaluated in terms of Full Width Half Maximum (FWHM), in the FoCal-E pixel layers have been found to be on the scale millimetres in good agreements with simulations. The resolution of FoCal-H~(cf.~Fig.~\ref{fig:focalTb}, right) was found to be about 16\% at \qty{100}{GeV} decreasing to about 11\% at \qty{350}{GeV}, satisfying the physics needs.

\section{Inner Tracking System 3 (ITS3)}\label{sec:ITS3}
\begin{figure}
  \centering
  \includegraphics[width=0.45\textwidth]{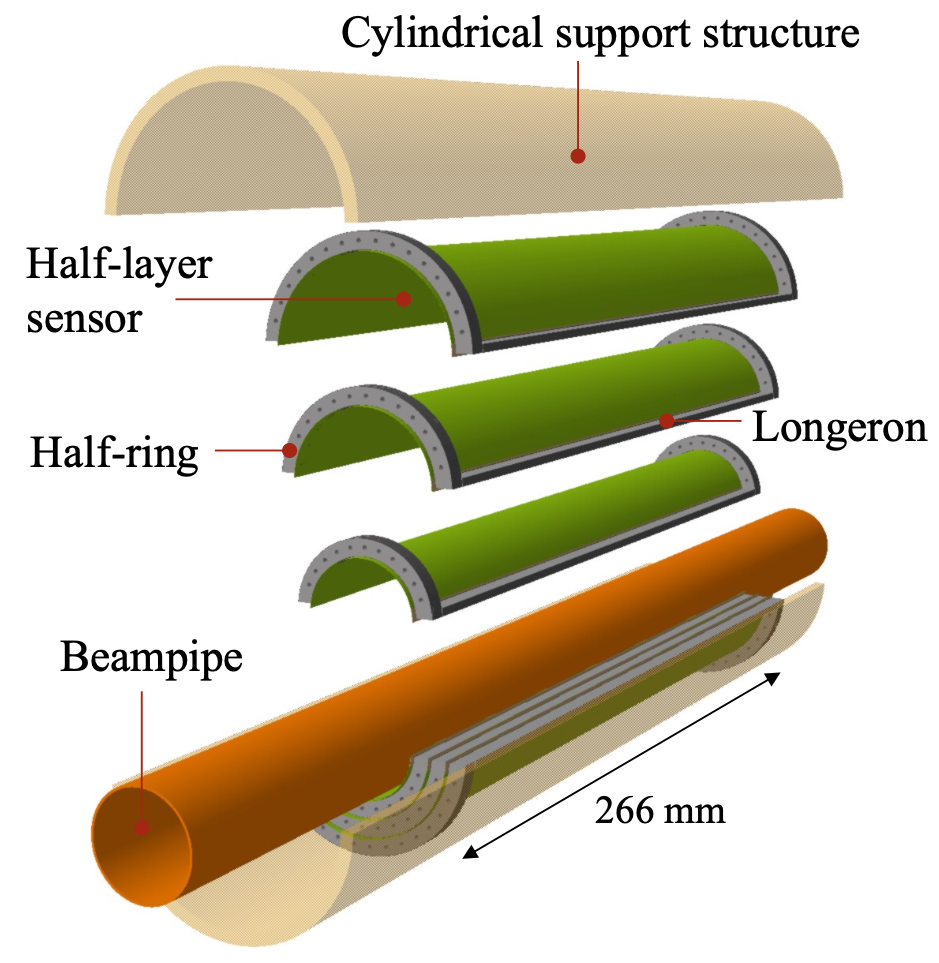}
  \hspace{0.5cm}
  \includegraphics[width=0.45\textwidth]{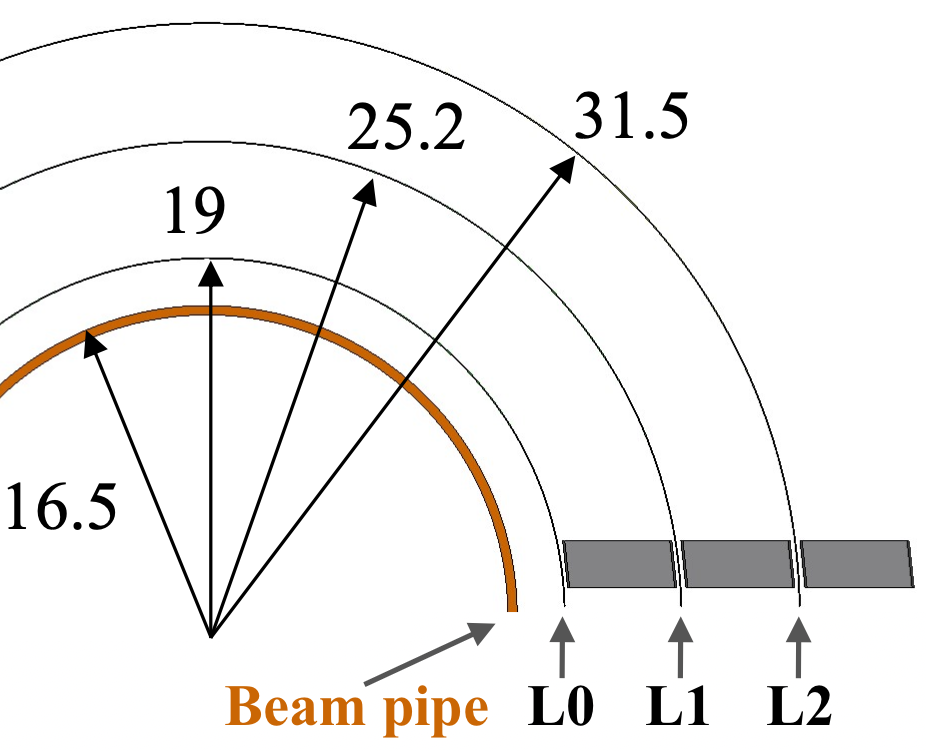}
  \caption{\textbf{Left:} schematic view of ITS3; \textbf{right:} \(r\varphi\)-view of ITS3 and the beam line, taken from~\cite{CERN-LHCC-2024-003}.\label{fig:ITS3}}
\end{figure}

\begin{figure}
  \centering
  \includegraphics[height=5cm]{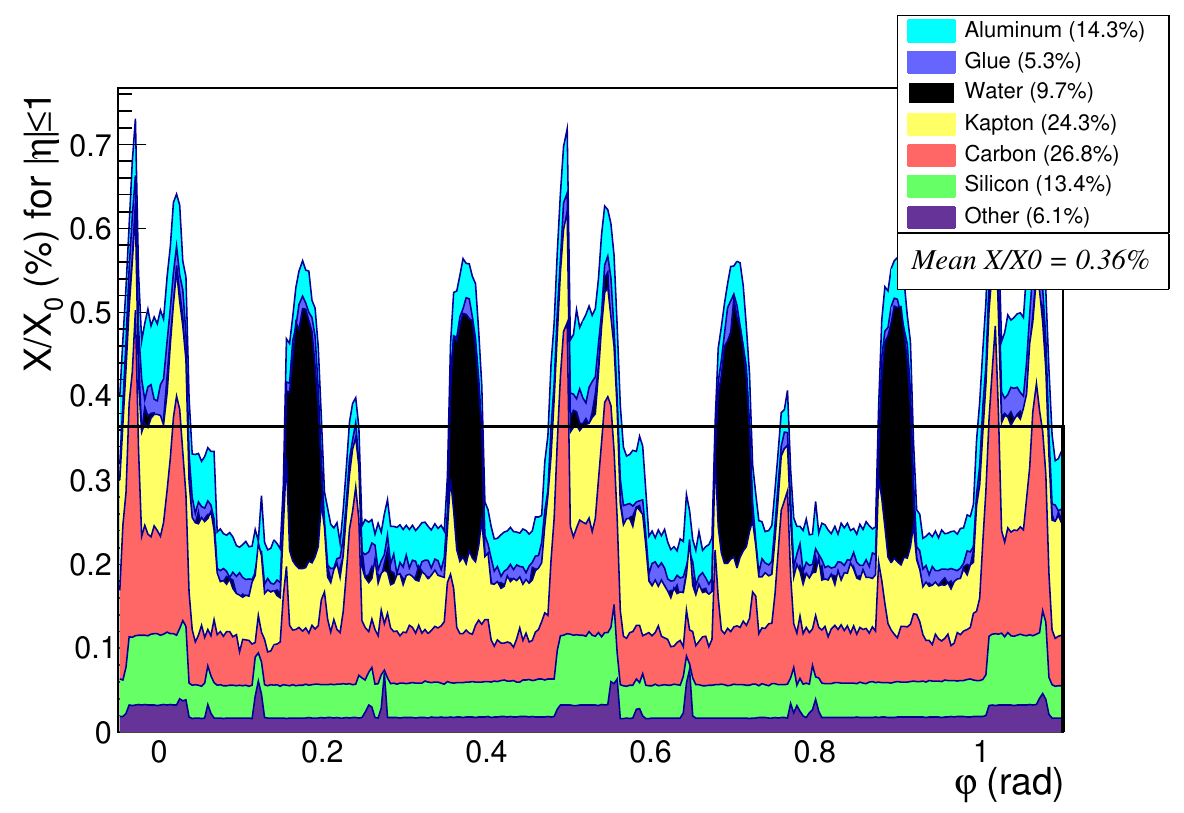}
  \hspace{0.5cm}
  \includegraphics[height=5cm]{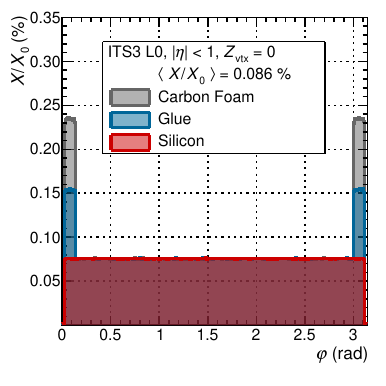}
  \caption{Material budget composition of the first layer for the current ITS (left) and ITS3 (right) as a function of \(\varphi\), taken from~\cite{Acharya_2024} and~\cite{CERN-LHCC-2024-004}, respectively.\label{fig:ITS3MB}}
\end{figure}

The Inner Tracking System 3 (ITS3)~\cite{CERN-LHCC-2024-003} will replace the innermost layers of the current Inner Tracking System 2 by truly cylindrical layers made from bent silicon achieving both a reduction of material budget and of the distance to the interaction point. As depicted in the schematic view in Fig~\ref{fig:ITS3}, there will be only little support material, namely the longerons and half-rings made from carbon foam. This is possible as every half-layer will be made of a single bent, large pixel sensor giving assembly mechanical stability. In addition, the detector will be air cooled, thus resulting in a reduction of the material budget per layer at midrapidity from 0.36\%~\(X_0\) to 0.09\%~\(X_0\) (cf.~Fig.~\ref{fig:ITS3MB}). Furthermore, the material budget distribution is very homogeneous and most of the acceptance is at low value of 0.07\%~\(X_0\)/layer. The reduction of the wall thickness of beam pipe from \qtyrange{800}{500}{\micro\meter} will also decrease the material between the interaction point and the first reconstructed hit point.

\begin{figure}
  \centering
  \includegraphics[width=0.7\textwidth]{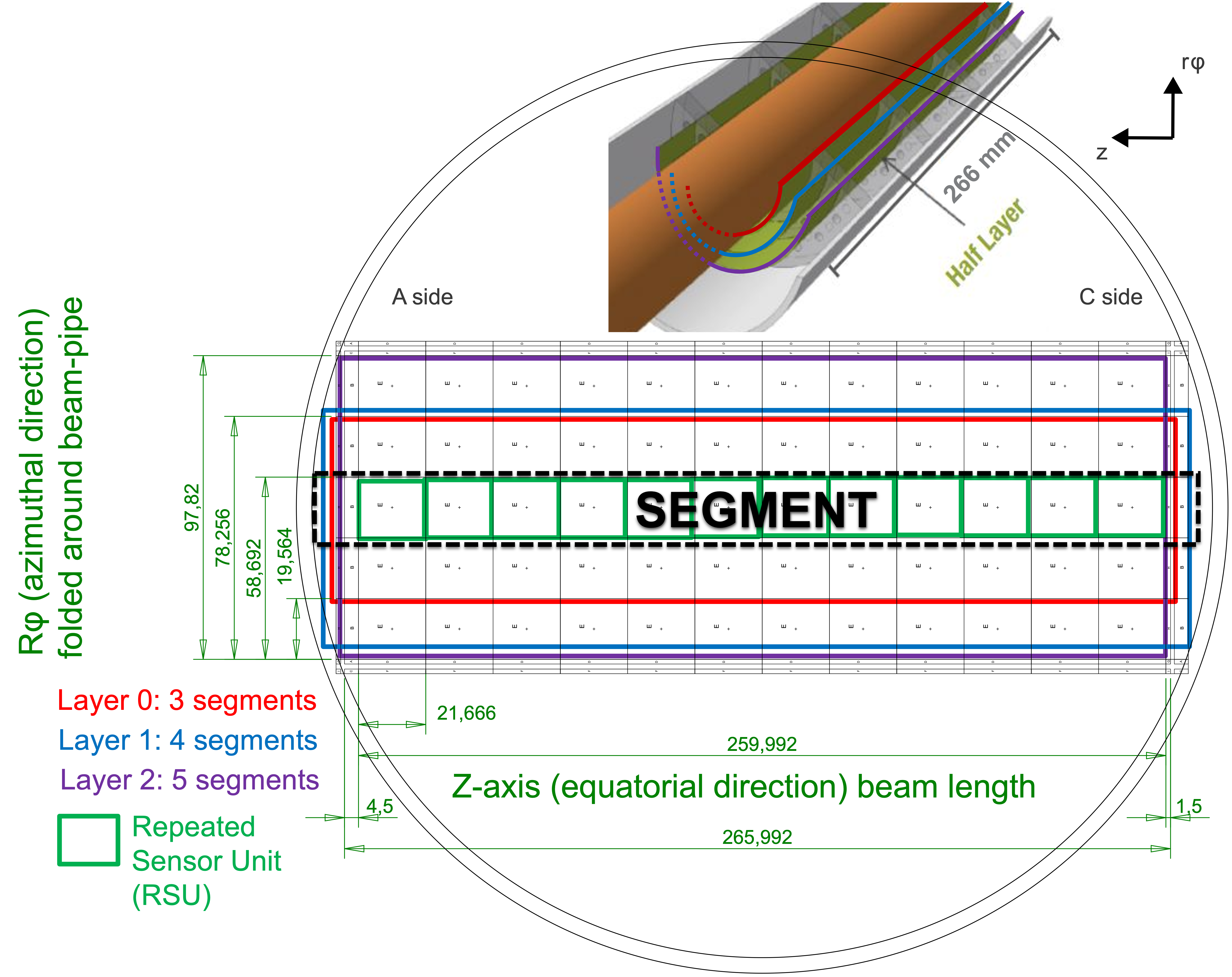}
  \caption{Schematic drawing of a MOSAIX wafer and the corresponding chip sizes employed to form the ITS3 half-layers.~\label{fig:ITS3MOSAIX}}
\end{figure}

The key element of ITS3 are wafer-scale pixel sensors based on stitched Monolithic Active Pixel Sensors (MAPS). These MAPS are manufactured in the TPSCo \SI{65}{nm} process. A first submission of stitched MAPS has demonstrated the feasibility of the concept and the MOSS prototype chip is operational and reaches full efficiency~\cite{CERN-LHCC-2024-003}. The next step is the submission of full size, full functionality sensor called MOSAIX. A MOSAIX segment will be made from twelve Repeated Sensor Units (RSUs), which in turn will comprise twelve tiles. Every tile can be powered independently, giving a segment 144-fold power granularity allowing to exclude possibly faulty portions of the chip. A full ITS3 Half-Layer will be built from \numrange{3}{5} layers (cf.~Fig~\ref{fig:ITS3MOSAIX}). To make this possible, MOSAIX will be interfaced from the extremities in beam direction. The submission of MOSAIX is foreseen for the beginning of 2025.

In terms of physics performance, reducing material budget and distance from the interaction point will lead to an improvement of the impact parameter resolution by about a factor of two, down to \qtyrange{10}{15}{GeV/\it{c}}. Many fundamental observables will strongly profit from this improvement or will become in reach. A summary of the ITS3 physics performance can be found in~\cite{ALICE-PUBLIC-2023-002}. For example, the measurement of \(\Lambda_b\) will become accessible in Pb--Pb collisions down to a transverse momentum of \qty{1}{GeV\per\textit{c}}, and will cover, with a statistical uncertainty smaller than 10\%, the region with the highest sensitivity to distinguish different quark coalescence models. Another example is the nuclear modification factor \(R_\mathrm{AA}\) for \(\mathrm{B}^{0}_s\) which in terms of precision is expected to allow to constrain model predictions for non-strange and strange B mesons.

\begin{figure}
  \centering
  \includegraphics[width=0.513\textwidth]{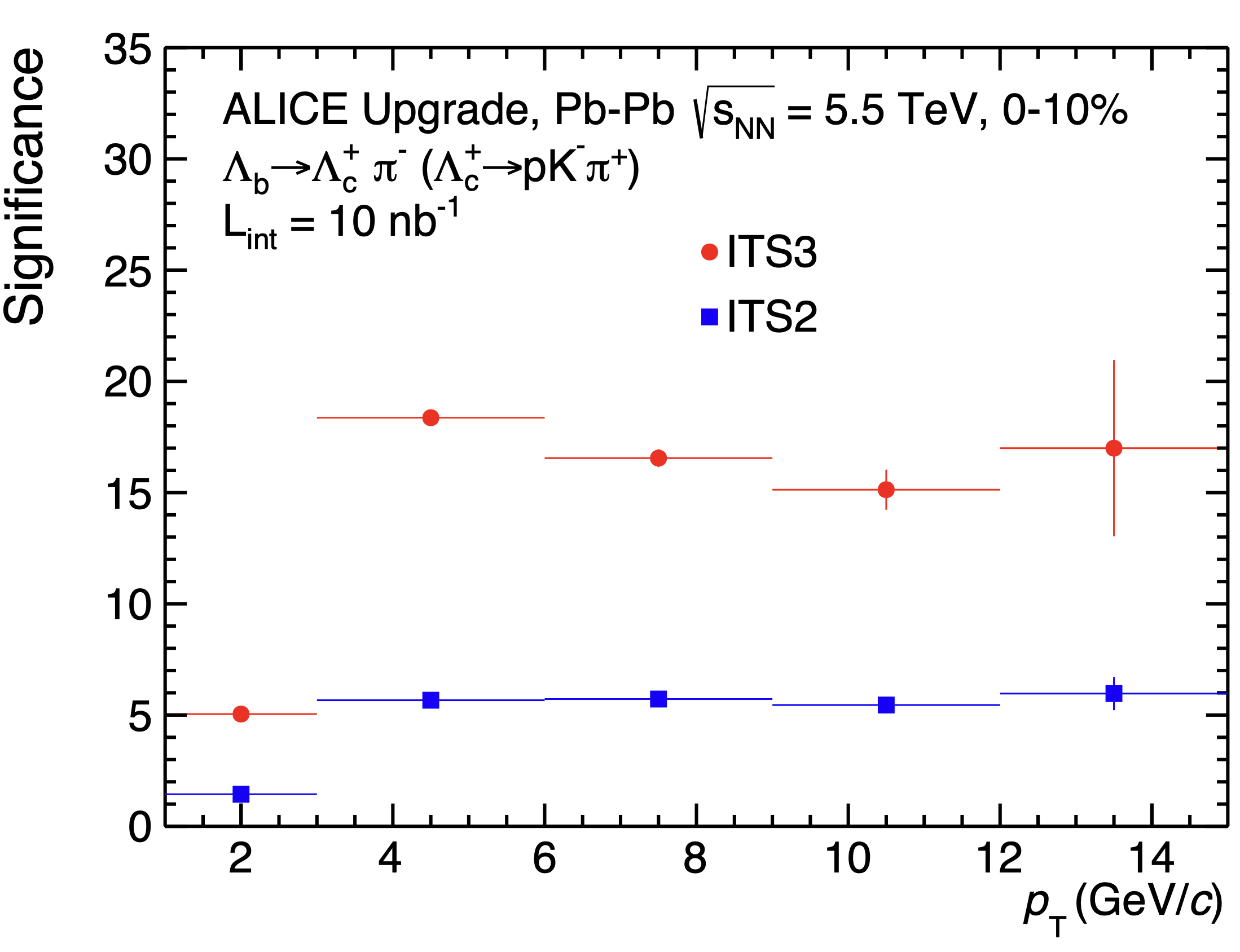}
  \hspace{0.5cm}
  \includegraphics[width=0.4\textwidth]{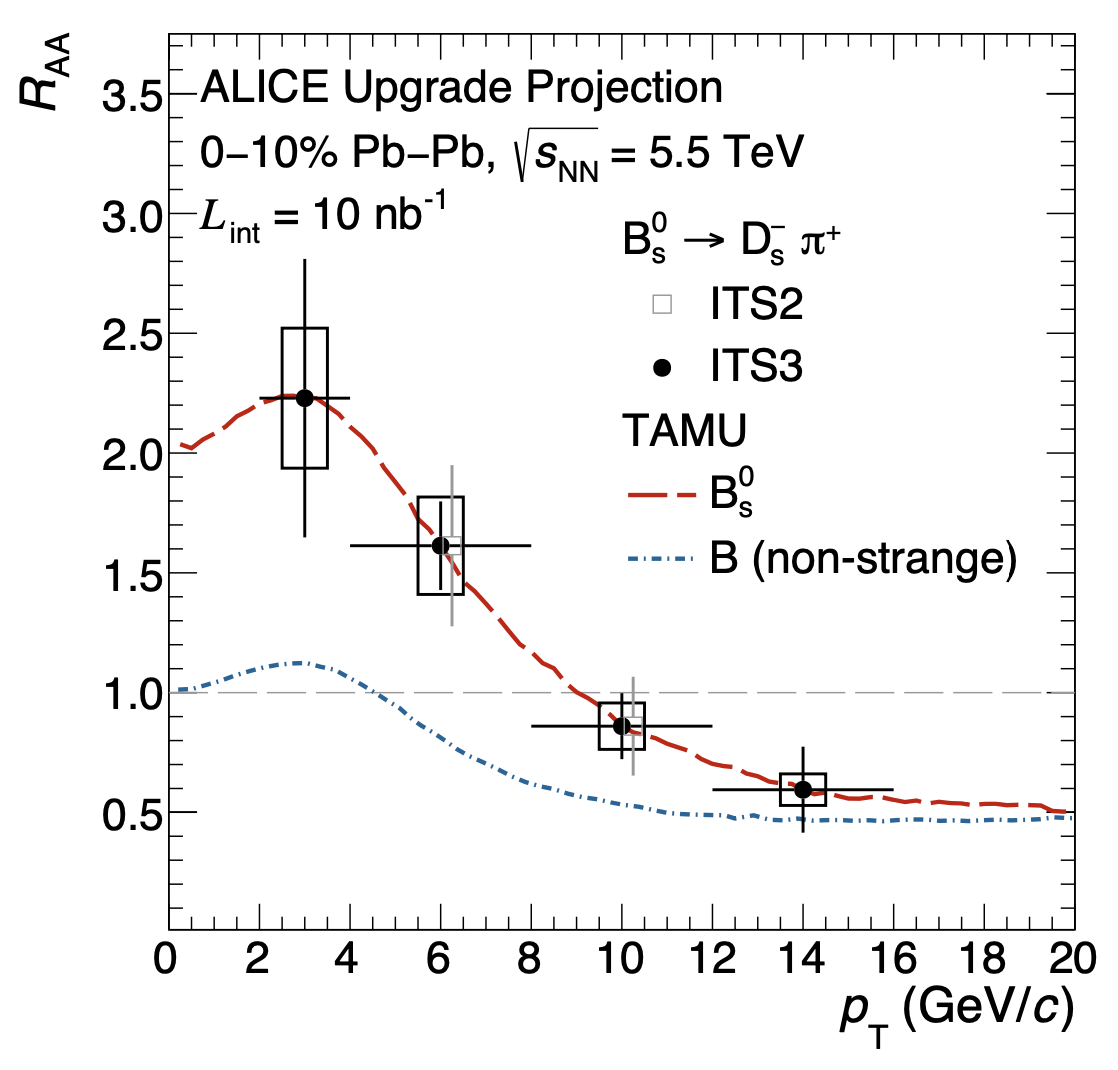}
  \caption{\textbf{Left:} projected significance for the full reconstruction of \(\Lambda_b\)-baryons in central Pb--Pb collisions; \textbf{right}: nuclear modification factor \(R_\mathrm{AA}\) for \(\mathrm{B}^{0}_s\), both taken from~\cite{ALICE-PUBLIC-2023-002}.\label{fig:ITS3physics}}
\end{figure}

\section{ALICE 3}\label{sec:ALICE3}
\begin{figure}
  \includegraphics[width=0.44\textwidth]{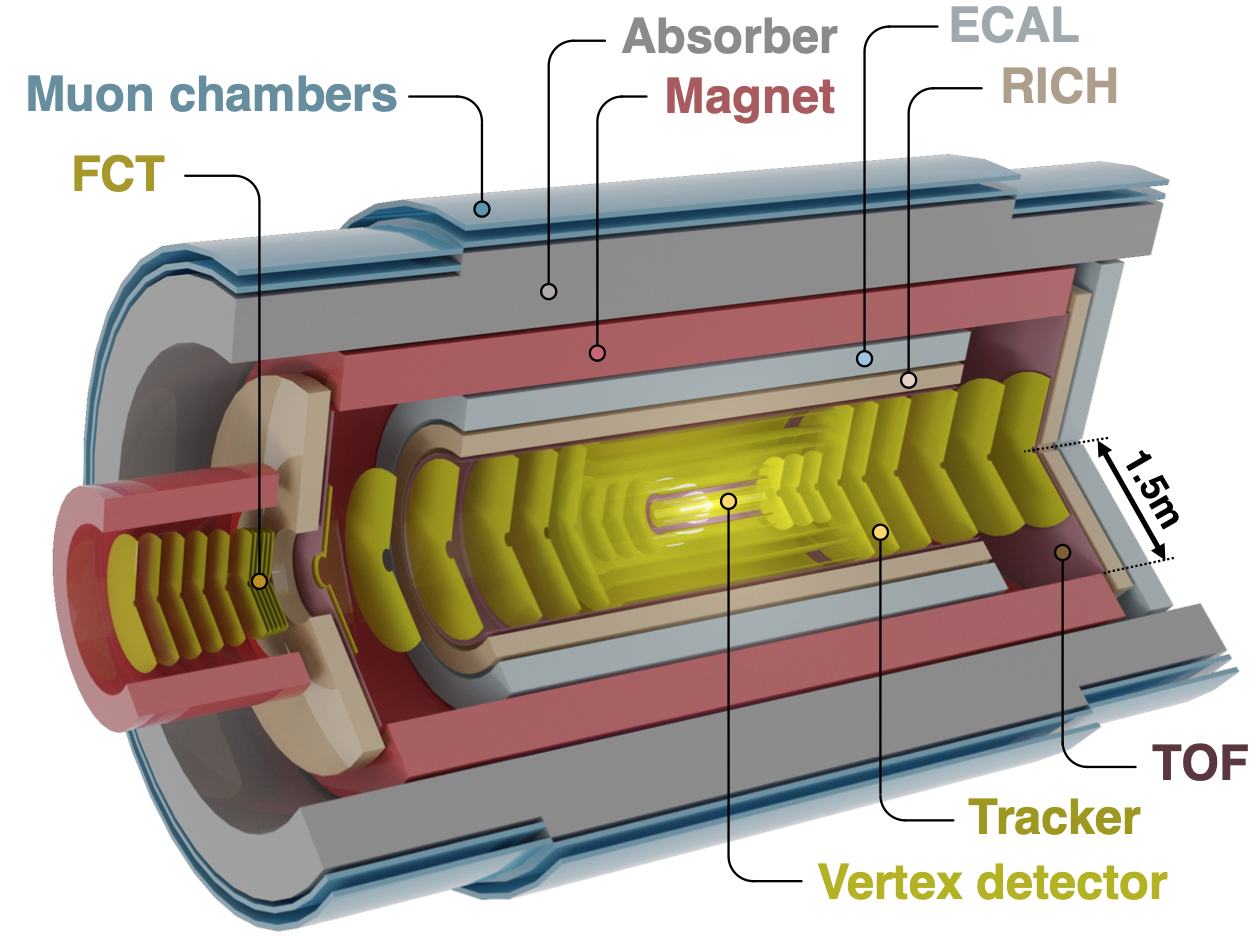}
  \hspace{0.3cm}
  \includegraphics[width=0.51\textwidth]{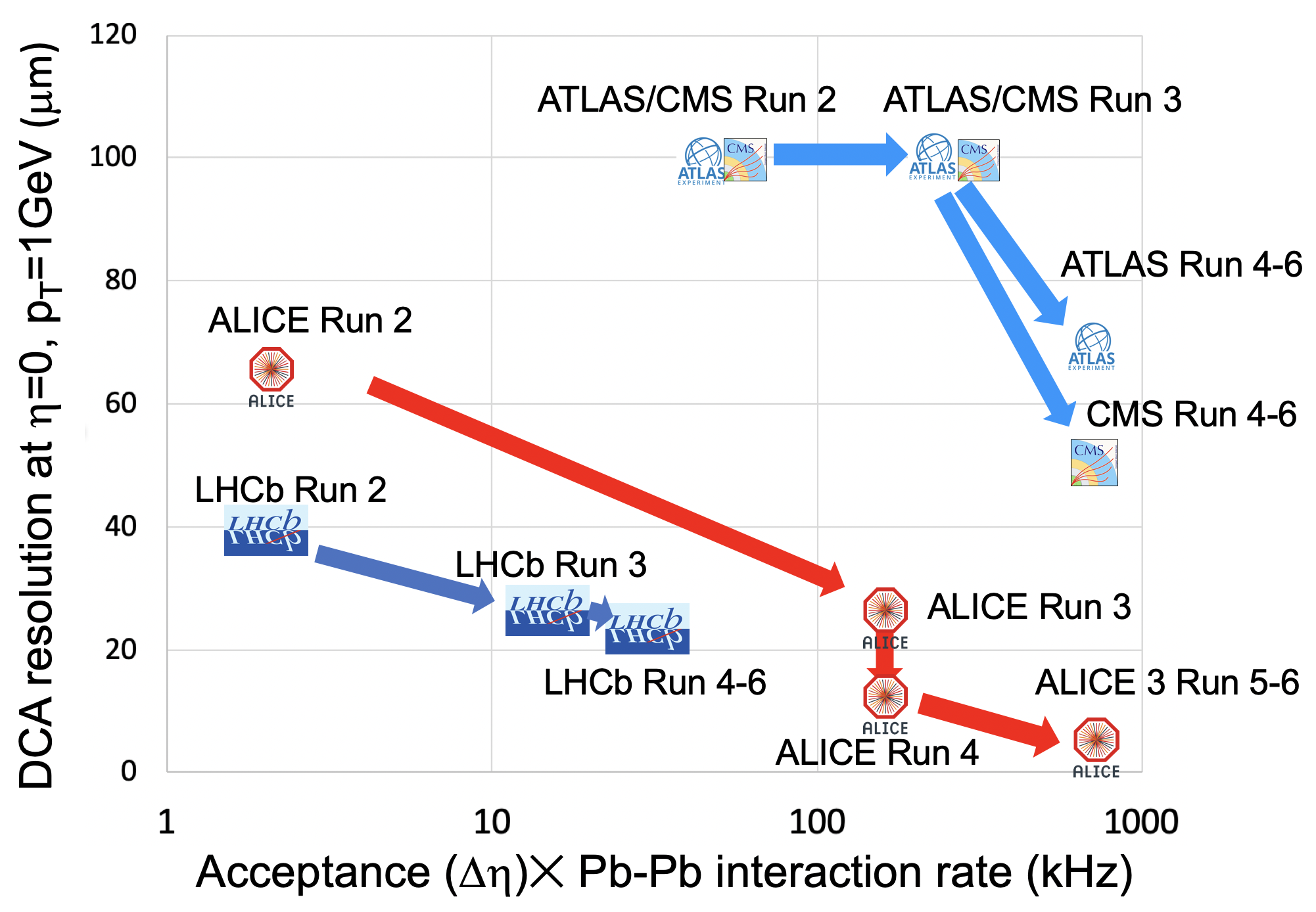}
  \caption{\textbf{Left:} sketch of the ALICE~3 detector, taken from~\cite{ALICE:2022wwr}; \textbf{right}: evolution of impact parameter resolution and acceptance times Pb--Pb interaction rate for the LHC experiments and their upgrades; ALICE, ATLAS and CMS resolutions quoted at midrapidity, LHCb resolutions quoted at \(\eta=3.5\).\label{fig:ALICE3}}
\end{figure}

ALICE~3~\cite{ALICE:2022wwr}~(cf.~Fig.~\ref{fig:ALICE3}, left.) is a novel and innovative detector concept for compact, large acceptance, low-mass all-silicon tracker. It will feature excellent vertex reconstruction combined with Particle IDentification (PID) capabilities. The ALICE~3 detector will be housed in a superconducting magnet in the range of \qtyrange{1}{2}{T}. ALICE~3 will further improve the impact parameter resolution as well as the Pb--Pb interaction rate capability~(cf.~Fig.~\ref{fig:ALICE3}, right) aiming to record pp collisions at \qty{24}{MHz} and acquire data samples of about \qty{20}{fb^{-1}} and \qty{35}{nb^{-1}} in pp and Pb--Pb, respectively.

\begin{figure}
  \includegraphics[width=0.42\textwidth]{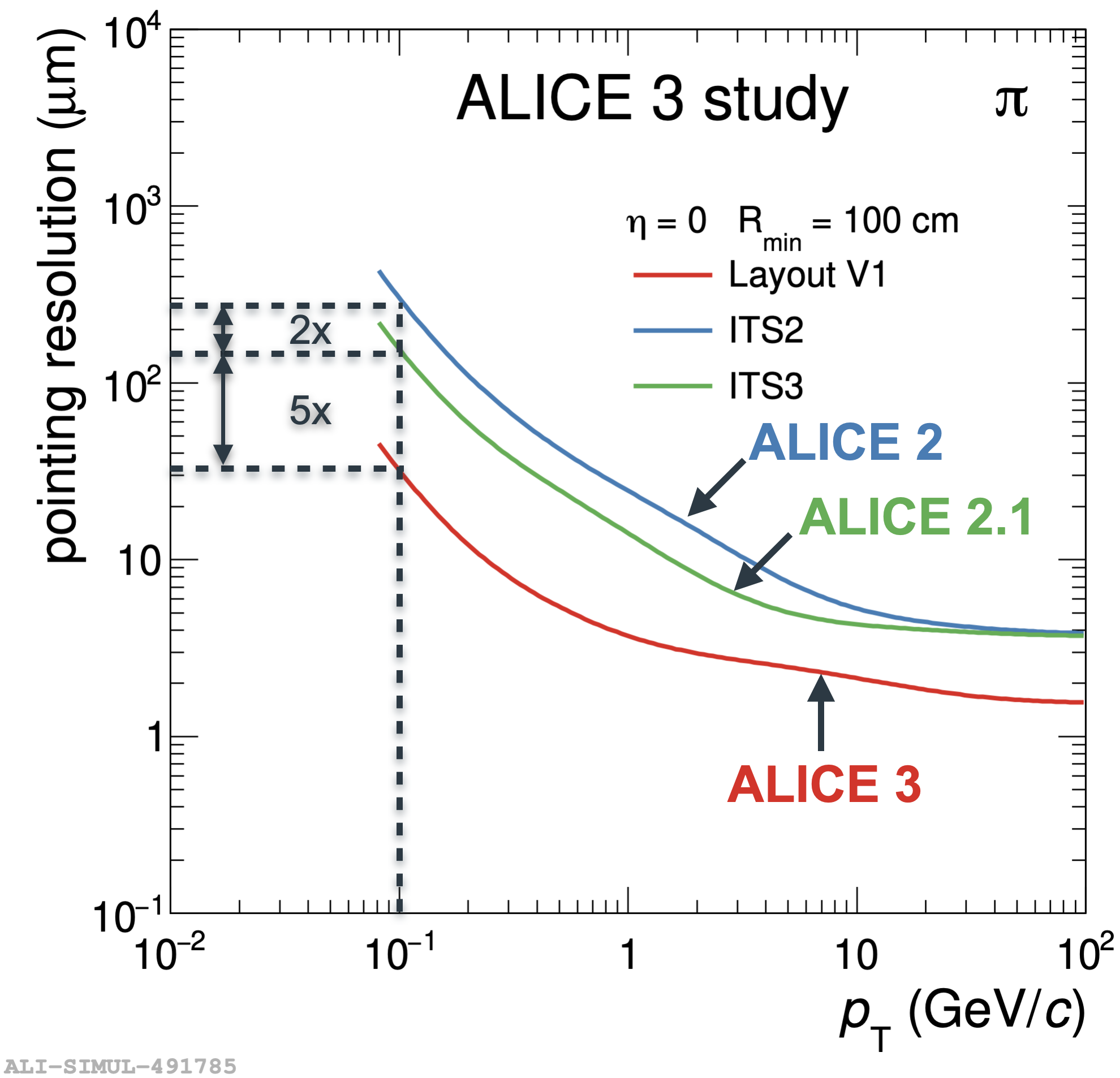}
  \includegraphics[width=0.53\textwidth]{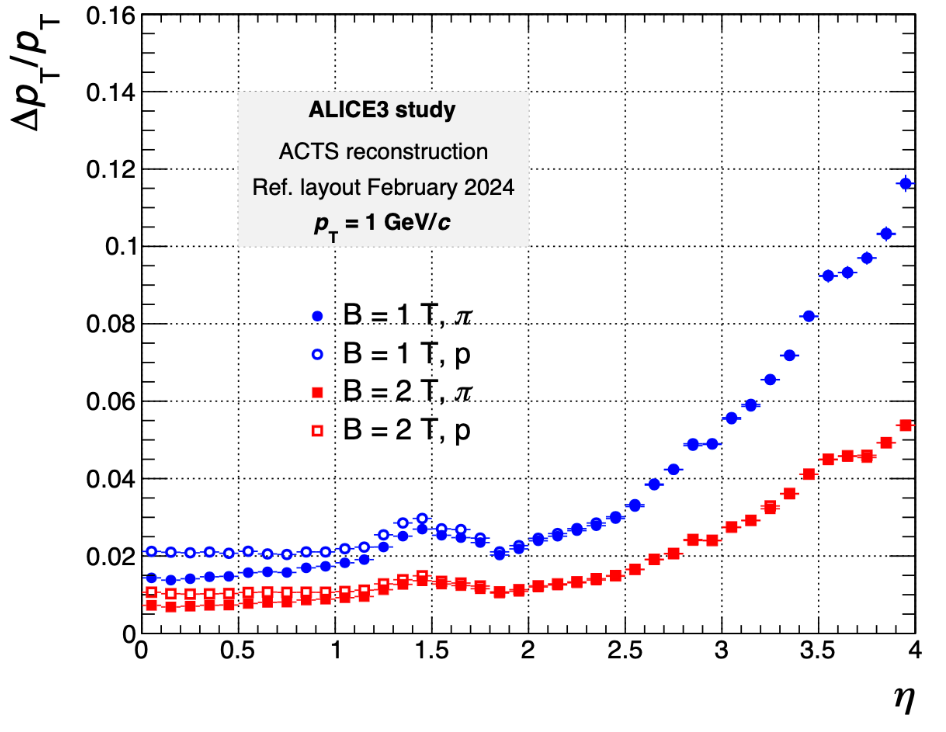}
  \caption{\textbf{Left:} projected point resolution for ALICE~3 for pions in comparison with the current ITS2 and the ITS3 upgrade; \textbf{right:} projected relative momentum resolution for ALICE~3 as function of pseudorapity for magnetic fields of \qty{1}{T} (blue) and \qty{2}{T} (red).\label{fig:ALICE3vtxTrk}}
\end{figure}

They key element for the improved impact parameter resolution at low transverse momentum is the Vertex Detector (VD). The minimal distance of the VD is given by the aperture of the beam which depends on the beam energy and is \qty{5}{mm} and \qty{15}{mm} at top energy and injection energy, respectively. Hence, a retractable vertex detector is required to approach the interaction point as close as possible at top energy. Combined with a low material budget of 0.1\%~\(X_0\)/layer and intrinsic spatial resolution of \qty{2.5}{\micro\metre}, the projected pointing resolution will be a factor 5 better than the one of ITS3 at a transverse momentum \SI{100}{MeV/\it{c}} (cf.~Fig.~\ref{fig:ALICE3vtxTrk}, left). The targeted small distance to the interaction point as well as the small material budget can only be achieved by placing the detector in a secondary vacuum inside the principle beam pipe. The required light-weight, in-vacuum mechanics and cooling pose a key R\&D challenge for the VD. As a consequence of the small distance from the interaction point as well as the targeted interaction rates, the hit-rate will amount to approximately \qty{100}{MHz\per\centi\metre\squared} and in turn the radiation load of the innermost layer will amount to \qty{1e16}{1\,MeV~n_{eq} \per\centi\metre\squared} and \qty{300}{Mrad}, one of the main challenges for MAPS R\&D for the VD. In the presence of such a radiation environment, the spatial resolution can only be achieved by a small pixel pitch of roughly \qty{10}{\micro\metre}, which in turn leads to a challenging integration of high-rate capable readout circuit and the in-pixel front-end circuit on a small surface. The VD will consist of three layers ranging from \qtyrange{5}{25}{\milli\metre} as well as three disks on either side.

The tracking layers of ALICE~3 will span radii ranging from \qtyrange{3.5}{80}{\centi\metre} for the eight barrel layers which will be accompanied by nine forwards disks per side to achieve a pseudorapidity coverage of \(\pm 4\). The total detector surface amounts to about \qty{60}{\metre\squared} making module design which allows for a high-yield, industrial mass production a key R\&D item. The material budget per layer will be 1\%~\(X_0\) and the spatial resolution \qty{10}{\micro\metre}, allowing for pixel pitches up to approximately \SI{50}{\micro\metre}. In order to achieve the targeted material budget, the services of the detector have to be kept to a minimum leading to a strict requirement of \qty{20}{mW\per\centi\metre\squared}. Limiting pile-up during pp data-taking requires a time resolution of \(O(\qty{100}{ns})\). In combination the requirements on power consumption and time resolution pose a key R\&D challenge.
The resulting tracking resolution is projected to be of the order of a few percent at mid-rapidity increasing by a factor of about five in the pseudorapidity range from~\numrange{2}{4} (cf.~Fig.~\ref{fig:ALICE3vtxTrk}, right).

\begin{figure}
  \includegraphics[width=0.41\textwidth]{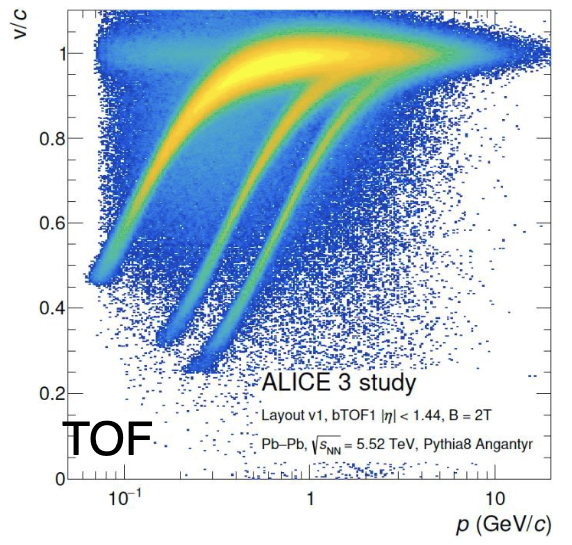}
  \includegraphics[width=0.58\textwidth]{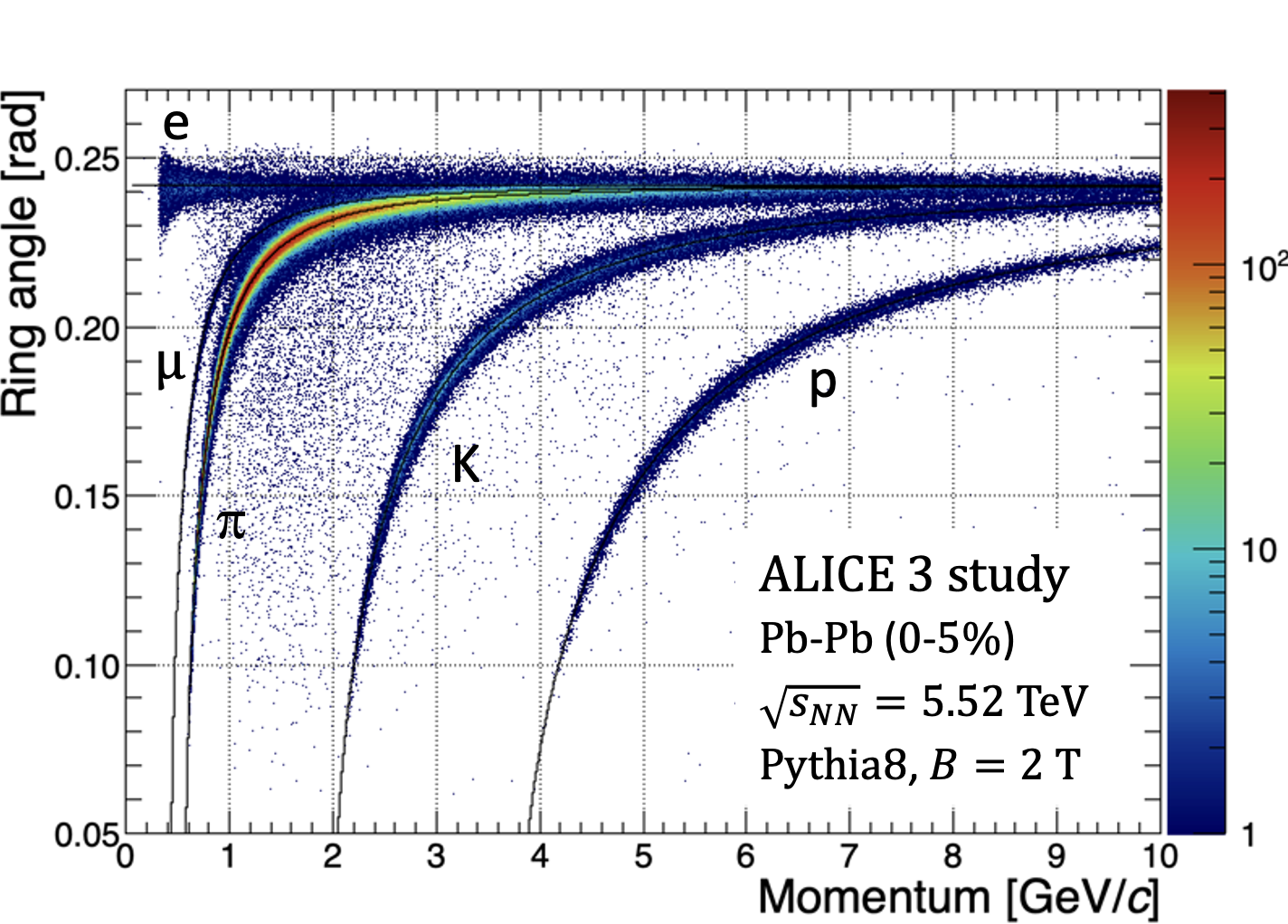}
  \caption{\textbf{Left:} projected velocity measurement performance of the ALICE~3 TOF; \textbf{right:} projected Cherenkov ring angle measurement for the ALICE~3 RICH.\label{fig:ALICE3PID}}
\end{figure}

The particle identification for ALICE 3 will be based on a combination of a Time-of-Flight (TOF) and a Ring-Imaging Cherenkov (RICH) detector. The TOF and RICH cover the lower and higher transverse momentum as depicted in Fig.~\ref{fig:ALICE3PID}.\\
For the TOF a targeted time resolution is \qty{20}{ps}. ALICE~3 is foreseen to feature two TOF layers, an inner TOF at roughly \qty{20}{\centi\metre} of radius with in the tracker and an outer TOF including disks right outside the tracker resulting in a total surface of about \SI{45}{\metre\squared}. The material budget per layer is required to be below 3\%~\(X_0\). There are currently three R\&D streams: Silicon Photon Multipliers (SiPMs) coated with different thicknesses and types of resins~\cite{ALICE3TOF2}, single and double Low Gain Avalanche Diodes (LGADs)~\cite{ALICE3TOF1} and CMOS-LGADs~\cite{Follo_2024}.
The RICH will be based on SiPMs and Aerogel radiator with refractive indices adapted to the momentum of the particles of \num{1.03} and \num{1.006} in the barrel and forward region, respectively. The total detector surface amounts to about \qty{35}{\metre\squared}. The main challenge for the R\&D is to achieve the radiation hardness required at the innermost disk radii.
Recently, also a combined detector based on a common photosensitive sensor performing both TOF and RICH measurements using a thin glass slab acting as Cherenkov radiator for TOF measurements has been proposed~\cite{10164558}.

In addition, ALICE~3 will also feature a Muon Identification system (MID) at central rapidity based on a steel hadron absorber and as baseline plastic scintillators read out via wave-length shifting fibres and SiPMs. Scintillators and MWPCs have been characterized at test beam to assess their performance in ALICE~3~\cite{Alfaro_2024}. The ALICE~3 MID is optimized for the reconstruction of charmonia down to zero transverse momentum.
Moreover, an electromagnetic calorimeter with a 2\(\pi\) coverage has been proposed. The sampling calorimeter will be based on about a hundred layers of \qty{1}{\milli\metre} Pb combined with \qty{1.5}{\milli\metre} scintillators. A PbWO\(_{4}\) segment will allow for high energy-resolution measurements.
The Forward Conversion Tracker (FCT) covering the pseudo rapidity region of $4<\eta<5$ will consist of very thin tracking disks located in a dedicated dipole magnet aiming to measure very low transverse momentum, below \qty{10}{MeV\per\it{c}}, photons. 

\begin{figure}
\includegraphics[width=0.44\textwidth]{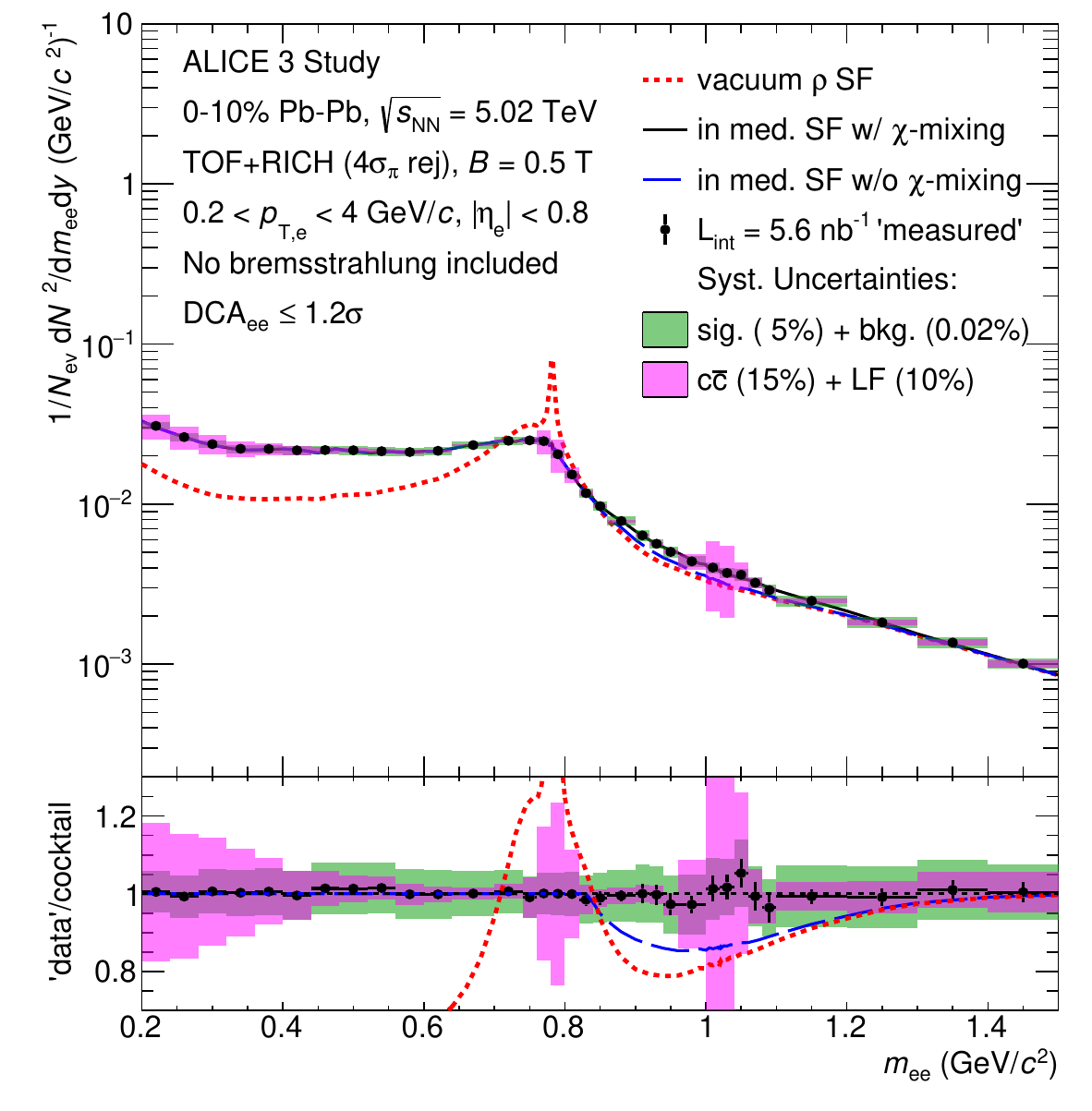}
\includegraphics[width=0.54\textwidth]{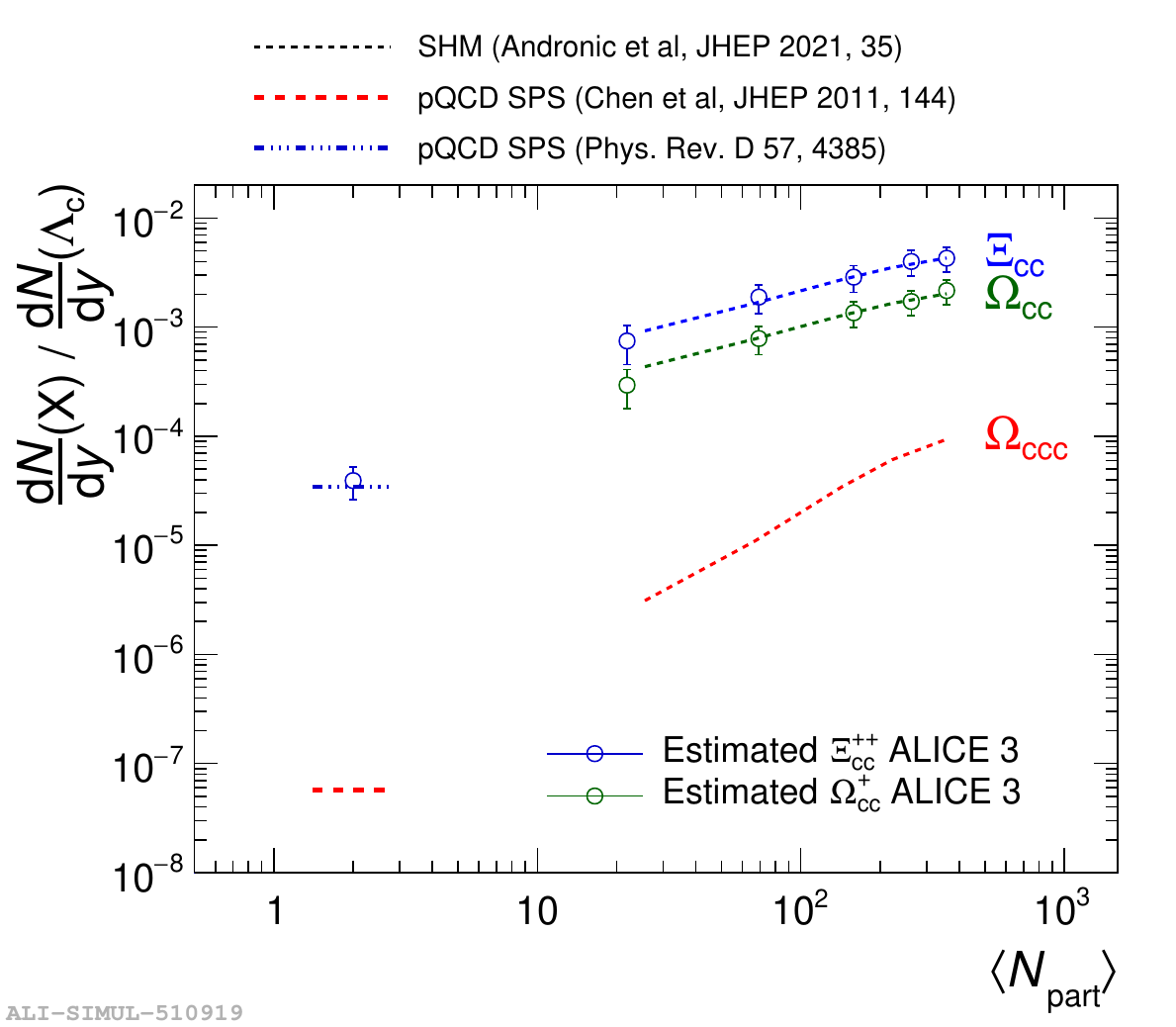}
\caption{\textbf{Left:} simulated dielectron spectrum in Pb--Pb collisions after subtraction of correlated light-hadron and heavy-flavour hadron decays, taken from~\cite{ALICE:2022wwr}; \textbf{right:} projected ratio of multi-charm (\(\Xi_{cc}\), \(\Omega_{cc}\), \(\Omega_{ccc}\)) to baryon to single charm baryon yield as a function of centrality in Pb--Pb collisions and in pp collisions, based on~\cite{ALICE:2022wwr}.\label{fig:ALICE3physics} }
\end{figure}

ALICE~3 targets to measure thermal dileptons with high precision to understand the time evolution of the QGP temperature as well as the mechanisms of chiral symmetry restoration. Figure~\ref{fig:ALICE3physics} (left) shows the expected measured thermal dilepton yield in comparison with calculation using the vacuum \(\rho\) spectral function (dashed red line) and medium-modified spectral function from a hadronic many-body approach with (black line) and without (dashed blue line) chiral mixing. With the projected precision, the ALICE 3 measurement will make the observation of effects due to chiral mixing possible.
With ALICE~3, it will become feasible to study the hadronisation of final state partons with multi-charm baryons. Multi-charm baryons require the combination of multiple independently produced charm quarks and hence are a unique probe for hadron formation. The Statistical Hadronisation Model~\cite{Andronic2021} predicts very large enhancement as well as a characteristic relation between yields of the different n-charm states. Figure~\ref{fig:ALICE3physics} (right) shows the expected performance for the measurement of the ratio of multi-charm (\(\Xi_{cc}\), \(\Omega_{cc}\), \(\Omega_{ccc}\)) to baryon to single charm baryon yield as a function of centrality in Pb--Pb collisions and in pp collisions.
An overview of the full ALICE~3 physics programme, which also includes heavy-flavour transport, exotica, nuclei, hadron-hadron interaction potentials, net-baryon fluctuations and ultrasoft photons can be found in~\cite{ALICE:2022wwr}.

\section{Summary and outlook}
ALICE has an ambitious upgrade program targeting to further the understanding of the QGP, in particular with precise measurements of heavy flavour and electromagnetic radiation. \\
The Run 4 upgrades are approaching the construction phase. FoCal will measure photons, neutral mesons and jets in the forward region to constrain nuclear gluon PDFs at low \(x\). ITS3 will be based on ultra-thin, truly cylindrical wafer-scale MAPS significantly improving secondary vertex reconstruction. \\
Beyond Run 4, ALICE~3 will fully exploit the high-luminosity LHC as heavy-ion collider throughout Run 6. For the foreseen novel, silicon-based detector concept, ALICE will be pioneering several R\&D directions impacting also other future high-energy physics experiments. The ALICE~3 apparatus will enable precision measurements of dileptons, multi-heavy-flavour hadrons and hadron correlations.

\bibliographystyle{JHEP}
\bibliography{literature}
\thispagestyle{empty}

\end{document}